\documentclass[onecolumn]{sdl}

\def\0{\mbox{\tiny $0$}}
\def\1{\mbox{\tiny $1$}}
\def\2{\mbox{\tiny $2$}}
\def\3{\mbox{\tiny $3$}}
\def\4{\mbox{\tiny $4$}}
\def\5{\mbox{\tiny $5$}}
\def\6{\mbox{\tiny $6$}}
\def\7{\mbox{\tiny $7$}}
\def\8{\mbox{\tiny $8$}}
\def\9{\mbox{\tiny $9$}}

\def\br{\mathbf{r}}

\def\p{\mbox{\tiny $\perp$}}
\def\s{\mbox{\tiny $\parallel$}}

\def\rel{\mbox{\tiny rel}}

\def\zR{z_{\mbox{\tiny R}}}
\def\w{\mathrm{w}_{\0}}

\def\g{\mbox{\tiny geo}}

\def\max{\mbox{\tiny max}}
\def\pol{\mbox{\tiny pol}}
\def\block{\mbox{\tiny block}}

\logo{
\colorbox{DarkGoldenrod}{\color{white}$\mathbf{\Sigma\hspace*{0.06cm} \delta \hspace*{0.04cm} \Lambda}$}
}

\journal{\shadowtext{\textbf{\color{DarkRed} European Physical Journal D}} \, \textbf{73}, 213-11 (2019)}

\titlelines{2}
\title{Power oscillations induced by\\ the  relative Goos-H\"anchen  phase}

\imgbgabstract{By using an optical interferometer composed of  a dielectric  laser  ellipsometer, to change the optical response of transverse electric and magnetic incident radiation, and two polarisers, to trigger  the interference pattern induced by the relative Goos-H\"anchen phase, we show under which conditions it is possible to optimize the laser power oscillations induced by the relative  phase difference between orthogonal polarised states. The  Goos-H\"anchen interference can then be used to sense rotation, to test optical components, and to simulate quarter and  half wave plates.}

\author{
\names{Manoel P. Ara\'ujo$^{1}$, Stefano De Leo$^{2,a}$, Gabriel G. Maia$^{3}$, and Maurizio Martino$^{4}$}
\affiliation{$^{1}$Institute of Physics Gleb Wataghin, State University of Campinas, Brazil}
\affiliation{$^{2}$Department of Applied Mathematics, State University of Campinas, Brazil}
\affiliation{$^{3}$Institute for Scientific and Industrial Research, University of Osaka, Japan}
\affiliation{$^{4}$Department of Mathematics and Physics, University of Salento, Italy}
 \email{$^{a}$deleo@ime.unicamp.br}
}

\begin{document}

\sdlmaketitle

\section{Introduction}

Interference is the hallmark of oscillatory phenomena \cite{book1,book2,book3,neutrino}. The most iconic support of this statement is perhaps the explanation interference provides for the fringe patterns in the Young's double slit experiment\cite{Young}, which settled the 19th century debate on the nature of light in favour of the oscillatory model. This experiment belongs to a general class of double path experiments, in which a wave is split into two separate waves that are later recombined into a single one. The difference in the optical paths results in a phase shift, creating an interference pattern. The relevance of this phenomenon can also be seen in the large body of techniques which came to be called collectively as Interferometry\cite{book4}. The guiding principle behind such techniques is simple: Interference is the net result of a superposition of waves travelling through the same medium and the phase difference between these waves generates an intensity pattern of which information about the waves can be drawn. The only constraint on this effect is that interference cannot happen between waves oscillating in {\em orthogonal} directions. This is true for a direct superposition but it does not mean a relative phase difference between orthogonal polarisation states is ineffective.  It is indeed possible, as we shall see in detail later, to induce an interference-like effect between orthogonal states using polarisers. In this case, they assume the role that weak interactions play in the production of Kaons\cite{book3} and neutrinos\cite{neutrino}.

The mixing between orthogonal states is also the basis of the optical analogue of weak measurements\cite{WM1,WM2,WM2b}. Nevertheless, in the optical weak measurement approach, the interest is often addressed to measure the Goos-H\"anchen (GH) shift \cite{GH1,ART,GH2}, being the GH relative phase removed by  waveplates \cite{WM3,WM4}. In this paper, we show how the interference between orthogonal polarised states can be triggered by an appropriate choice of the polarisers set-up and how the relative GH phase between the orthogonal polarised states induces power oscillations in the optical beam which, in the regime of total internal reflection, is transmitted through a dielectric block.

Let us begin by considering light coming out of a laser source. It has a definite polarisation state and its electric field can be described as
\begin{equation}
\mathcal{E}(\br) = E(\br)\,\left[\begin{array}{cc}
a_{\s} \\ a_{\p}
\end{array}\right]\,\,,
\end{equation}
where
 \[
E(\br)=   E_{\0} \,\frac{\w^{\2}}{4\,\pi}\int \hspace*{-0.1cm} \mbox{d}k_x\,\mbox{d}k_y\,\,\,\exp\left[\,-\,\frac{(\,k_x^{\2}+k_y^{\2}\,)\,\w^{\2}}{4 }\,+\,i\,{\mathbf{k}} \cdot {\mathbf{r}}\,\right] =
\frac{E_{\0}\,e^{ikz}}{1+i\,z/\zR}\,
\exp\left[\,-\,\frac{x^{\2}+y^{\2}}{\w^{\2}\,(\, 1+i\,z/\zR \,)}\,\right]
\]
is the laser profile, $z$ the coordinate parallel to the light' propagation direction,
 $\zR=\pi\w^{\2}/\lambda$ the Rayleigh length, $\lambda$ the beam wavelength, $k=2\,\pi/\lambda$ the wave number, and $\w$ the minimal distribution waist. The paraxial approximation, $k_z=\sqrt{k^{\2}-k_x^{\2}-k_y^{\2}}\approx k - (k_x^{\2}+k_y^{\2})/2\,k$, allows the analytical integration and the condition
 $a_{\s}^{\2}+a_{\p}^{\2}=1$
 sets the field normalisation. After interacting with a polariser set at an angle $\alpha$, see Fig.\,1(a), the resultant field is modified by the Jones matrix associated to the polariser,
\begin{equation}
J_{\alpha} = \left(\begin{array}{cc}
\cos^{\2}\alpha & \sin\alpha\cos\alpha \\ \sin\alpha\cos\alpha & \sin^{\2}\alpha
\end{array}\right)\,\,,
\end{equation}
as follows
\begin{equation}{\label{eq:Ealpha}}
\mathcal{E}_{\alpha}(\mathbf{r}) =  J_{\alpha} \, \mathcal{E}(\br)
=  \left( a_{\s} \,\cos\alpha \,+\, a_{\p} \,\sin\alpha \right)\, E(\br)\,\left[\begin{array}{c}
\cos\alpha \\ \sin\alpha
\end{array}\right]\,.
\end{equation}
If there is a relative phase between the components of the original field, the polariser produces a complex amplitude which will exhibit an interference pattern based on the nature of the relative phase. Usually, relative phases are induced by a differentiation of optical paths by means of beam splitters and mirrors, but to study this orthogonal interference phenomenon we will focus on an optical system which presents responses intrinsically different to transverse electric (TE) and transverse magnetic (TM) polarised light: The total internal reflection regime inside a dielectric block. When light is totally internally reflected the Fresnel coefficient associated with the reflection becomes complex and the electric field acquires a polarisation-dependent phase known as GH phase. As a consequence, the propagation direction of the field is laterally shifted, a phenomenon known as GH shift\cite{GH1,ART,GH2}. This shift has been the matter of investigations \cite{GHS1,GHS2,GHS3,GHS4,GHS5,GHS6,GHS7,GHS8} for decades. It is proportional to the wavelength of the light being shifted and so its detection normally involves amplification techniques such as weak measurements \cite{WM1,WM2,WM2b,WM3,WM4}.
In this case, the used light beam  is a composite of TE and TM polarisations. In weak measurements the relative GH phase between TE and TM components is removed. In this paper, we do not remove this relative phase and we mix orthogonal polarisation states by an appropriate choice of the second polarization angle in order to produce a particular intensity pattern detected by a power measurement. From the GH interference fringes information about the GH shift and optical components can be indirectly obtained.

The paper is organised as follows: In the next section, we describe the optical system of interest as well as the incoming and transmitted beams. Section III is devoted to obtain the expression for the power oscillation and section IV discusses the relevant aspects of the theory and possible experimental implementations. Our final thoughts are presented in the final section.

\section{The Goos-H\"anchen interferometer}

Let us consider a system composed of a dielectric block with right angles as depicted in Fig.\,1(a). The coordinate system $x_{*}yz_{*}$ follows the geometry of the block with the $x_{*}$ coordinate placed along its height and the $z_{*}$ coordinate along it length, see Fig.\,1(b). We will also define a coordinate system $xyz$ with $z$ along the propagation direction of the incident beam and $x$ perpendicular to $z$ but still contained in the optical plane, see Fig.\,1(b). Light coming out of the laser source passes through a polariser set at an angle $\alpha$, acquiring the form described in
Eq.\,({\ref{eq:Ealpha}}), entering then the dielectric block by its left face making an angle $\theta$ with its normal,
\begin{equation}
\left(\begin{array}{c} x_* \\ z_* \end{array} \right)= \left( \begin{array}{rr} \cos\theta & \sin\theta\\ -\,\sin\theta & \cos\theta\end{array}\,\right)\, \left(\begin{array}{c} x \\ z \end{array} \right)\,\,,
\end{equation}
 see Fig\,1(b). The transmission angle, $\psi$, is given by the Snell law, $\sin\theta= n\,\sin\psi$ and the incidence and reflection angles at the bottom and top faces of the dielectric, $\varphi$, by the geometry of the system, $\varphi = \frac{\pi}{2}-\psi$.

A theoretical approach may simply state the number of times light will be totally internally reflected without concern for how to obtain such reflections in a real laboratory. From an experimental perspective, however, the capabilities of the optical systems at hand must be known. It can be shown that, for the dielectric block we are considering, the relation
\begin{equation}{\label{eq:LAB}}
\overline{AD} = 2\,\tan\varphi\,\,\overline{AB} =  2\,\frac{\sqrt{n^{\2}-\sin^{\2}\theta}}{\sin\theta}\,\,\overline{AB}\,\,,
\end{equation}
where $\overline{AD}$ is the length of the block and $\overline{AB}$ its height, guarantees two total internal reflections for a beam entering the block at an incidence angle $\theta$, with the entrance point being at half the height $\overline{AB}$. For a block of given dimensions, Eq. (\ref{eq:LAB}) tells us at which angle the beam, after two internal reflections, touches the right face of the block at the same height in which it touches the left one.
By replacing the unitary block with $N$ blocks with lengths that are a multiple of the original length, we ensure $2\,N$ reflections and the power can be measured as a function of the optical system  size or equivalently as a function of the number of internal reflections. In this article, we shall consider two possibilities, i.e. building the blocks with acrilic or using a container of water whose length can be varied by using a mechanical system, see Fig.\,1(a). If the walls of the container are thin walls they do not contribute significantly to the light transmission so the results presented in this paper for dielectric water are valid in any laboratory system in which a container with thin walls is used.

Once the beam comes out of the dielectric, its original distribution is modified by the Fresnel transmission coefficient of the optical system. We shall suppose that the minimal waist is at the point in which the incident beam touches the left face of the  dielectric block. For the mathematical notation, this means setting the origin of the cartesian axes
in this point, see Fig.\,1(b). In the $x_*yz_*$ system, representing the reflection/transmission coefficient system, the
wave number vectors of the incident, left transmitted, down reflected, and down transmitted  beams will be respectively $(k_{x_*},k_{y},k_{z_*})$, $(k_{x_*},k_{y},q_{z_*})$, and $(p_{x_*},p_{y},q_{z_*})$, with  $q_{z_*}^{\2}-k_{z_*}^{\2}=(n^{\2}-1)\,k^{\2}$ and  $p_{x_*}^{\2}-k_{x_*}^{\2}=(1-n^{\2})\,k^{\2}$. The  transmission coefficient, given in terms of these wave numbers, is
obtained by multiplying the transmission coefficient at the left air/dielectric interface, the double (internal) reflection coefficient at the down and up dielectric/air interfaces, and finally the  transmission at the right dielectric/air interface.

The transmission coefficient at the left air/dielectric interface is given by
\begin{equation}
T_{\pol}^{^{\mathrm{[left]}}}(k_x,k_y) = \frac{2\,\sigma_{\pol}\,k_{z_*}}{\sigma_{\pol}\,k_{z_*}+\,q_{z_*}}\,\,,
\end{equation}
with $(\sigma_{\s},\sigma_{\p})=(n^{\2},1)$. The double internal reflection for a dielectric structure composed of $N$ unitary blocks of dimensions (\ref{eq:LAB}) leads to the following coefficient
\begin{equation}
R^{^{\mathrm{[int]}}}_{\pol}(k_x,k_y) = \left(\frac{k_{x_*}-\sigma_{\pol}\,p_{x_*}}{k_{x_*}+\,\sigma_{\pol}\,p_{x_*}}\right)^{2\,N}\,\exp[\,2\,i\,N\,k_{x_{*}}\,
\overline{AB}\,]\,\,.
\end{equation}
The transmission coefficient at the right dielectric/air interface is given by
\begin{equation}
T^{^{\mathrm{[right]}}}_{\pol}(k_x,k_y) = \frac{2\,q_{z_*}}{q_{z_*}+\,\sigma_{\pol}\,k_{z_*}}\,\,\exp[\,i\,N\,(q_{z_*}-\,k_{z_*})\,\overline{AD}\,]\,\,.
\end{equation}
In the regime of total internal reflection, $n\sin\varphi>1$ or equivalently $\sin\theta<\sqrt{n^{\2}-1}$, the wave number  $p_{x_*}$ becomes imaginary, $p_{x_*}=i\,|p_{x_*}|=i\,\sqrt{(n^{\2}-1)k^{\2}-k_{x_*}^{\2}}$, and the wave number distribution, describing the outgoing transmitted beam is given by
\begin{eqnarray}
\label{WND1}
G_{\pol}(k_x,k_y;\br) & = &   \frac{4\,\sigma_{\pol}\,k_{z_*}q_{z_*}}{(\sigma_{\pol}\,k_{z_*}+\,q_{z_*})^{^{2}}}\,\,g(k_x,k_y;\br)\,\,\times \nonumber \\
 & & \exp\left\{\,i\,\left[\,N\,(q_{z_*}-\,k_{z_*})\,\overline{AD}\,+\,2\,\,N\,k_{x_{*}}\,
\overline{AB} \,-\,4\,N\,\arctan\left(\frac{\sigma_{\pol}\,|p_{x_*}|}{k_{x_*}}\right)
\right]\right\}\,\,,
\end{eqnarray}
where
$g(k_x,k_y;\br) = \exp\left[\,-\,(\,k_x^{\2}+k_y^{\2}\,)\,\w^{\2}/\,4\,+\,i\,{\mathbf{k}} \cdot {\mathbf{r}}\,\right]$
is the wave number distribution of the incoming gaussian beam.
In the paraxial approximation the beam is strongly collimated around the direction of incidence and so  phases in the previous equation  can be expanded up to first order around the center of the gaussian distribution $|g(k_x,k_y;\br)|$
 and the transmission coefficient taken at $(k_x,k_y)=(0,0)$. In this way, the wave number distribution given in (\ref{WND1}) becomes
\begin{equation}
\label{WND2}
G_{\pol}(k_x,k_y;\br)\,=\,\mathcal{A}_{\pol}\,\,g(k_x,k_y;\mathbf{r_{\pol}})\,\,
\exp[\,i\,(\,\phi+\phi_{\pol}\,)\,]\,\,,
\end{equation}
where
\begin{eqnarray*}
\mathcal{A}_{\pol} & = & 4\,\sigma_{\pol}\cos\theta\sqrt{n^{\2}-\sin^{\2}\theta}\,\,\mbox{\LARGE /}
\left(\sigma_{\pol}\,\cos\theta+\sqrt{n^{\2}-\sin^{\2}\theta}\,
\right)^{^{2}}\,\,,\\
\phi &  = &  k\,\left[\,N\,\left(\sqrt{n^{\2}-\sin^{\2}\theta}-\cos\theta\right)\,\overline{AD}\,+\,2\,N\,\sin\theta\,\overline{AB}\,\right]\,\,,\\
\phi_{\pol} & = &
-\,4\,N\,\arctan\left[\,\sigma_{\pol}\,\sqrt{n^{\2}-1-\sin^{\2}\theta}\,/\,\sin\theta\,\right]\,\,,
\end{eqnarray*}
and
\begin{eqnarray*}
\mathbf{r_{\pol}} &=& \left\{\,x+d+d_{\pol}\,,\,y\,,\,z\,\right\}\nonumber \\
& = &
\left\{\,x\,+\,N\,\sin\theta\,\overline{AD}\,+\,
\frac{4\,N\,\sigma_{\pol}\,(n^{\2}-1)\,\cos\theta}{k\,[\,(n^{\2}-1)\,\sigma_{\pol}^{\2}\,+\,(1-\sigma_{\pol}^{\2})\,\sin^{\2}\theta\,]\,
\sqrt{n^{\2}-1-\sin^{\2}\theta}}\,,\,y \,,\,z \,\right\}\,\,.
\end{eqnarray*}
The lateral shift $d$, proportional to $\overline{AD}$,  is the displacement predicted by the geometrical optics by using the Snell and reflection laws. The additional shift  $d_{\pol}$, proportional to $\lambda$, is the well-known GH shift\cite{GH1,ART,GH2}. The geometrical shift is the same for TE and TM waves, whereas the GH shift is polarisation dependent. This is a direct consequence of the fact that the geometrical phase only depends on the boundary conditions while, in the regime of total internal reflection, the polarisation phase derives from the  Fresnel reflection coefficient.

The approximation of the transmitted wave number distributions,  given in Eq.\,(\ref{WND2}), allows to obtain an analytical expression for the TE and TM outgoing beams,
\begin{equation}
\label{Epol}
E_{\pol}(\br)\,=\,\mathcal{A}_{\pol}\,\,E(\mathbf{r_{\pol}})\,\,
\exp[\,i\,(\,\phi +\phi_{\pol}\,)\,]\,\,.
\end{equation}

\section{The effects induced by the  Goos-H\"anchen phase}

The incoming beam $\mathcal{E}_{\alpha}(\mathbf{r})$, after passing through the dielectric block,
is modified to
\begin{equation}
\label{Eblock}
\mathcal{E}_{\block}(\br)\,=\,
\left( a_{\s} \,\cos\alpha \,+\, a_{\p} \,\sin\alpha \right)\, \left[\begin{array}{c}
\mathcal{A}_{\s} \,E(\mathbf{r_{\s}})\,e^{i\,\phi_{\s}}\cos\alpha \\
\mathcal{A}_{\p}\, E(\mathbf{r_{\p}})\,e^{i\,\phi_{\p}}\,\sin\alpha
\end{array}\right]\,\, e^{i\,\phi_{\g}}\,\,.
\end{equation}
A power measurement of this beam does not contain any oscillation phenomenon. In fact,
\begin{equation}
P_{\block} = \int \hspace*{-0.1cm} \mbox{d}x\,\mbox{d}y\,\,\left|\mathcal{E}_{\block}(\br)\right|^{^{2}}=
\left( a_{\s} \,\cos\alpha \,+\, a_{\p} \,\sin\alpha \right)^{^{2}}\,\left( \mathcal{A}_{\s}^{^{2}}\,\cos^{\2}\alpha\,+\, \mathcal{A}_{\p}^{^{2}}\,\sin^{\2}\alpha     \right)\,P_{\0}\,\,,
\end{equation}
where
\[ P_{\0}=\displaystyle{\int} \hspace*{-0.1cm} \mbox{d}x\,\mbox{d}y\,\,\left|E(\mathbf{r_{\s}})\right|^{^{2}} = \displaystyle{\int} \hspace*{-0.1cm} \mbox{d}x\,\mbox{d}y\,\,\left|E(\mathbf{r_{\p}})\right|^{^{2}} = \displaystyle{\int} \hspace*{-0.1cm} \mbox{d}x\,\mbox{d}y\,\,\left|E(\br)\right|^{^{2}} = \pi\,|E_{\0}|^{^{2}}\w^{\2}\,/\,2\,\,.
\]
To trigger  interference between the orthogonal polarised states, we use a second polariser at an angle $\beta$. After passing through it, the optical beam is then described by  $\mathcal{E}_{\beta}=J_{\beta}\,\mathcal{E}_{\block}$. Explicitly,
\begin{eqnarray}
\label{Ebeta}
\mathcal{E}_{\beta}(\br) &=&
\left( a_{\s} \,\cos\alpha \,+\, a_{\p} \,\sin\alpha \right)\,\times \nonumber \\
& & \left[
\mathcal{A}_{\s} \,E(\mathbf{r_{\s}})\,e^{{i\,\phi_{\s}}}\cos\alpha\cos\beta \,+\,
\mathcal{A}_{\p}\, E(\mathbf{r_{\p}})\,e^{{i\,\phi_{\p}}}\sin\alpha\sin\beta\right]\,e^{{i\,\phi}}\,
\left[\begin{array}{c}
\cos\beta \\
\sin\beta
\end{array}\right]\,\,,
\end{eqnarray}
and its intensity is given by
\begin{eqnarray}
\label{Ibeta}
\left|\mathcal{E}_{\beta}(\br)\right|^{^{2}} &=&
\left( a_{\s} \,\cos\alpha \,+\, a_{\p} \,\sin\alpha \right)^{^2}\,
\left\{\mathcal{A}_{\s}^{^{2}} \,\left|E(\mathbf{r_{\s}})\right|^{^{2}}\cos^{\2}\alpha\cos^{\2}\beta \,+\,
\mathcal{A}_{\p}^{^{2}} \,\left|E(\mathbf{r_{\p}})\right|^{^{2}}\sin^{\2}\alpha\sin^{\2}\beta \right. +\nonumber \\
 & & \hspace*{4.1cm} \left. \frac{\mathcal{A}_{\p} \,\mathcal{A}_{\s}\, \sin2\alpha\,\sin2 \beta }{2}\,
 \mbox{Re}\left[e^{{i\,(\phi_{\p}-\,\,\phi_{\s})}}\,E(\mathbf{r_{\p}})\,E^{^{*}}(\mathbf{r_{\s}}) \right] \right\}.
\end{eqnarray}
By observing that
\[
\int \hspace*{-0.1cm} \mbox{d}x\,\mbox{d}y\,\, E(\mathbf{r_{\p}})\,E^{^{*}}(\mathbf{r_{\s}}) = \exp\left[-\,\frac{(d_{\p}-d_{\s})^{^2}}{2\, \w^{\2}}\right]\,P_{\0}\,\,,
\]
we find for the beam power measured after the second polariser the following expression
\begin{eqnarray}
\label{Pbeta}
P_{\beta} &=& \int \hspace*{-0.1cm} \mbox{d}x\,\mbox{d}y\,\,\left|\mathcal{E}_{\beta}(\br)\right|^{^{2}}
\,=\, \left( a_{\s} \,\cos\alpha \,+\, a_{\p} \,\sin\alpha \right)^{^2}\,\times \\
& & \hspace*{-1.5cm}\,\left\{\mathcal{A}_{\s}^{^{2}}\cos^{\2}\alpha\cos^{\2}\beta \,+\,
\mathcal{A}_{\p}^{^{2}}\sin^{\2}\alpha\sin^{\2}\beta \,+\,\frac{\mathcal{A}_{\p} \,\mathcal{A}_{\s}\, \sin2\alpha\,\sin2 \beta }{2}\,\cos(\phi_{\p}-\,\,\phi_{\s})\,\exp\left[-\,\frac{(d_{\p}-d_{\s})^{^2}}{2\, \w^{\2}}\right]\right\}\,P_{\0}\,\,.\nonumber
\end{eqnarray}
The relative power $\mathcal{P}_{\rel}=P_{\beta}/P_{\block}$ is the given by
\begin{equation}
\label{relpow}
\mathcal{P}_{\rel}(\alpha,\beta;\tau) = \frac{2\,\tau^{\2}\cos^{\2}\alpha\cos^{\2}\beta \,+\,2\,
\sin^{\2}\alpha\sin^{\2}\beta \,+\,\tau\, \sin2\alpha\,\sin2 \beta \,\cos\Phi_{_{\mathrm{GH}}}\exp[-\,\delta^{^2}_{_{\mathrm{GH}}}/2]}{2\,(\tau^{\2}\cos^{\2}\alpha +\sin^{\2}\alpha)}\,\,,
\end{equation}
where
\[
\tau=\mathcal{A}_{\s}/\mathcal{A}_{\p}=   \left[\left(\cos\theta\,+\,\sqrt{n^{\2}-\sin^{\2}\theta}\,\right)\,\mbox{\LARGE $/$}\,\left(n\,\cos\theta\,+\,\sqrt{1-(\sin\theta/n)^{^2}}\,\right)\right]^{{\2}}\,\,,
\]
\[\delta_{_{\mathrm{GH}}}= \frac{d_{\p} - d_{\s}}{\mathrm{w}_{\0}} =4\,N\,\frac{[\,n^{\2}(n^{\2}-1)-(n^{\2}+1)\sin^{\2}\theta\,]\,\cos\theta}{
k\,{\mathrm{w}_{\0}}\,(\sin^{\2}\theta-n^{\2})\,\sqrt{n^{^{2}}-1-\sin^{\2}\theta}
}\,\,,\]
and
\[
\Phi_{_{\mathrm{GH}}}=\phi_{\p} -\phi_{\s} = 4\,N\,\arctan\left[\,\sqrt{n^{\2}-\sin^{\2}\theta-1}\,\,\sin\theta\,/\, (n^{\2}-\sin^{\2}\theta)\right]\,\,.
\]
In Fig.\,2, the relative GH phase, $\Phi_{_{\mathrm{GH}}}$, is plotted for a different number $N$ of unitary dielectric blocks made of acrylic and water (note that a water container of thin walls does not modify the results presented in this paper). The choice of these dielectrics is motivated by two reasons. The first one is their easy availability and the second one is the fact that the acrylic and water refractive indexes, $n= 1.490$ and $n=1.332$, are greater and lesser than $\sqrt{2}$. Remembering that the constraint for total internal reflection is $\sin\theta<\sqrt{n^{\2}-1}$, for a refractive index greater than $\sqrt{2}$, for all incidence angles we remain always in the regime of total internal reflection, see Fig.\,2(a), whereas for the water, an upper limit exists, i.e. $\theta_{_{\mathrm{w}}}=61.63^{\mathrm{o}}$, see Fig.\,2(b).

The maximum of the relative GH phase,
\begin{equation}
\label{GHmax}
\left\{\,\theta_{\max}\,,\,\,\Phi_{_{\mathrm{GH}}}(\theta_{\max})\,\right\} =  \left\{\,\arcsin\left(n\,\sqrt{\frac{n^{\2}-1}{n^{\2}+1}}\right)\,,\, 4\,N\,\arctan\left(\frac{n^{\2}-1}{2\,n}\right)  \,\right\}\,\,,
\end{equation}
is found at the angle for which  the GH lateral displacements for TM and TE are equal, $\delta_{_{\mathrm{GH}}}(\theta_{\max})=0$.
For acrylic and water, we have
\begin{equation}
\begin{array}{lcl}
\left\{\,\theta_{\max}\,,\,\,\Phi_{_{\mathrm{GH}}}(\theta_{\max})/N\,\right\}_{_{\mathrm{acrylic}}} &= &
\left\{\,66.52^{\mathrm{o}}\,,\,1.55\,\right\}\,\,,\\
\left\{\,\theta_{\max}\,,\,\,\Phi_{_{\mathrm{GH}}}(\theta_{\max})/N\,\right\}_{_{\mathrm{water}}} &= &
\left\{\,44.72^{\mathrm{o}}\,,\,1.13\,\right\}\,\,.
\end{array}
\end{equation}
The maximum value of the relative GH phase for a dielectric block of a given refractive index allows to determine the minimal $N$-blocks configuration able to produce a full cycle of power oscillation. In this spirit, it is also important to give the solutions of the equation $\Phi_{_{\mathrm{GH}}}=a\,\pi$,
\begin{equation}
\label{solpi}
\theta(a,N)=\arcsin\left[ \sqrt{n^{\2}-\frac{n^{\2}+1}{2}\,\cos^{\2}\left(\frac{a\,\pi}{4\,N}\right)\pm \sqrt{\left(\frac{n^{\2}+1}{2}\right)^{\2}\,\cos^{\4}\left(\frac{a\,\pi}{4\,N}\right) - n^{\2} \cos^{\2}\left(\frac{a\,\pi}{4\,N}\right) } }   \,\,  \right]\,\,.
\end{equation}

\section{Looking at experimental implementations}

In the previous section, we essentially studied two effects on the propagation of a laser beam. The first one is caused by the transmission through a dielectric block, built to guarantee $2\,N$ total internal reflections and it is  mathematically described by the relative GH phase. The second one  is caused by the second polariser found after the dielectric block which trigger the GH interference.

\subsection{Simulating quarter and half wave plates}

An interesting aspect of the GH phase is that right after the beam is transmitted through the dielectric block, the polarisation of the incoming beam is changed. This happens because each component of the electric field responds differently to the block. If the incidence angle and the number of reflections are chosen carefully, the TE and TM components are transmitted in equal proportions and the control of the relative GH phase allows  then circular polarisation. This is an elegant alternative to the wave plates, usually employed to this end, since it does not require birefringent materials, but moreover,   it allows a broader control over the retardation angle.

After coming out from the dielectric block and before reaching the second polariser, the electric field is proportional to
\[
\left[ \begin{array}{c} \tau\, \cos\alpha \\ \exp(\,i\,\Phi_{_{\mathrm{GH}}}\,)\,\sin\alpha \end{array} \right]\,\,.   \]
Observing that  the difference between the transmission coefficient for TM ($A_{\s}$) and TE ($A_{\p}$) waves is quite small for incidence angles smaller than  $\pi/4$,  we can use the relative GH phase to reproduce the effects of quarter and  half wave plates.  By choosing the appropriate combination of  dielectric blocks and incidence angles, we can, for example, introduce a unitary  imaginary complex factor  between the TM and TE components of the electric field, $\Phi_{\mathrm{GH}}=\pi/2$, generating circularly polarised light, simulating a quarter-wave plate. There are many possible configurations $(\theta,\overline{AD}/\mathrm{cm})_{_{N}}$ for achieving it. The incidence angle $\theta$ is  calculated by Eq.\,(\ref{solpi}) fixing $a=1/2$ and varying $N$. Once the incidence angle is obtained, we can determine the length of the unitary block $\overline{AD}$ by using Eq.\,(\ref{eq:LAB}). For $\overline{AB}=1\,$cm, the possible combinations $N\leq6$ are
\[
\left\{\,
(7.56^{\mathrm{o}},22.56)_{\6}\,\,,\,\,\,\,\,
(9.09^{\mathrm{o}},18.76)_{\5}\,\,,\,\,\,\,\,
(11.40^{\mathrm{o}},14.94)_{\4}\,\,,\,\,\,\,\,
(15.30^{\mathrm{o}},11.11)_{\3}\,\,,\,\,\,\,\,
(23.45^{\mathrm{o}},7.22)_{\2}
\,\right\}_{_{\mathrm{acrylic}}}
\]
for acrylic, see Fig.\,2(a), and
\[
\left\{\,
(7.61^{\mathrm{o}},20.02)_{\6}\,\,,\,\,\,\,\,
(9.15^{\mathrm{o}},16.63)_{\5}\,\,,\,\,\,\,\,
(11.50^{\mathrm{o}},13.21)_{\4}\,\,,\,\,\,\,\,
(15.51^{\mathrm{o}},9.76)_{\3}\,\,,\,\,\,\,\,
(24.23^{\mathrm{o}},6.18)_{\2}
\,\right\}_{_{\mathrm{water}}}\,\,.
\]
for water, see Fig.\,2(b).  In order to choose the best configuration, it is also important to know the $\tau$ factor for the corresponding angles. For acrylic, we have
\[
\left\{\,
(7.56^{\mathrm{o}},1.0019)\,\,,\,\,\,
(9.09^{\mathrm{o}},1.0028)\,\,,\,\,\,
(11.40^{\mathrm{o}},1.0044)\,\,,\,\,\,
(15.30^{\mathrm{o}},1.0080)\,\,,\,\,\,
(23.45^{\mathrm{o}},1.0195)
\,\right\}_{_{\tau,\,\mathrm{acrylic}}}
\]
and, for water,
\[
\left\{\,
(7.61^{\mathrm{o}},1.0011)\,\,,\,\,\,
(9.15^{\mathrm{o}},1.0016)\,\,,\,\,\,
(11.50^{\mathrm{o}},1.0026)\,\,,\,\,\,
(15.51^{\mathrm{o}},1.0047)\,\,,\,\,\,
(24.23^{\mathrm{o}},1.0121)
\,\right\}_{_{\tau,\,\mathrm{water}}}
\]
Half-wave plates generates a phase shift of $\pi$ between the polarization components. It is possible to reproduce the half-wave plate effect by using one of the following $(\theta,\overline{AD}/\mathrm{cm})_{_{N}}$     combinations
\[
\left\{\,
(15.30^{\mathrm{o}},11.11)_{\6}\,\,,\,\,\,\,\,
(18.50^{\mathrm{o}},9.18)_{\5}\,\,,\,\,\,\,\,
(23.45^{\mathrm{o}},7.22)_{\4}\,\,,\,\,\,\,\,
(32.43^{\mathrm{o}},5.18)_{\3}
\,\right\}_{_{\mathrm{acrylic}}},
\]
for acrylic, and
\[
\left\{\,
(15.51^{\mathrm{o}},9.76)_{\6}\,\,,\,\,\,\,\,
(18.85^{\mathrm{o}},8.00)_{\5}\,\,,\,\,\,\,\,
(24.23^{\mathrm{o}},6.18)_{\4}\,\,,\,\,\,\,\,
(35.68^{\mathrm{o}},4.11)_{\3}
\,\right\}_{_{\mathrm{water}}}\,\,,
\]
for water.  See Fig.\,2(a) and (b) respectively. The $\tau$ factors are
\[
\left\{\,
(15.30^{\mathrm{o}},1.0080)\,\,,\,\,\,
(18.50^{\mathrm{o}},1.0118)\,\,,\,\,\,
(23.45^{\mathrm{o}},1.0195)\,\,,\,\,\,
(32.43^{\mathrm{o}},1.0402)
\,\right\}_{_{\tau,\,\mathrm{acrylic}}}
\]
for acrylic and
\[
\left\{\,
(15.51^{\mathrm{o}},1.0047)\,\,,\,\,\,
(18.85^{\mathrm{o}},1.0071)\,\,,\,\,\,
(24.23^{\mathrm{o}},1.0121)\,\,,\,\,\,
(35.68^{\mathrm{o}},1.0293)
\,\right\}_{_{\tau,\,\mathrm{water}}}
\]
for water. Note that the $\overline{AD}$ given in the previous expressions represents the length  of the unitary block for which the incoming and the outgoing laser beam have the same height for the incident (left side) and exit  (right side) point. This allows two internal reflections for a dielectric structure builded by using $N$ unitary blocks. It is interesting to observe that a dielectric block of length between $N\,\overline{AD} - \overline{AD}/4$  and $N\,\overline{AD}+\overline{AD}/4$ will continue to guarantee $2\,N$ internal reflections. Thus, by using the previous results,  a laser beam composed by an $\alpha$ mixture of TM and TE waves incident on an acrylic block of length $\in (2\times 7.22 - 7.22/4, 2\times 7.22 +7.22/4)$\,cm$= (12.635,16.245)$\,cm, and forming an angle $23.45^{\mathrm{o}}$ with the normal at the left air/dielectric interface of the block, after propagation gains a circular polarisation and this optical system simulates a quarter-wave plate. By using a dielectric block of length $\in (3\times 5.18 - 5.18/4, 3\times 5.18 +5.18/4)$\,cm$= (14.245,16.835)$\,cm and choosing as incidence angle $32.43^{\mathrm{o}}$, after propagation, we simulate a half-wave plate. This shows that an acrylic dielectric block
of length $15$\,cm and height $1$\,cm can be used to simulate a quarter-wave plate for incidence at  $23.45^{\mathrm{o}}$ and an half-wave plate for incidence at $32.43^{\mathrm{o}}$.  We can repeat what was done for the acrylic and find for water in a (thin walls) container of length $12$\,cm that an incidence at $24.23^{\mathrm{o}}$
and $35.60^{\mathrm{o}}$ simulate a quarter and half-wave plate respectively.

Dielectric blocks, when used to produce elliptically polarised light, are called Fresnel Rhomb retarders. They behave as wave plates, with the exception that they are made of non-birefringent materials. The descriptions of this device found in literature usually consider a normal incidence and the constraint of two or more internal reflections.
To simulate a quarter-wave plate by using  a  Fresnel rhomb, the acrylic and water dielectric blocks  have to be built in order to guarantee at least three internal reflections. The angle, $\gamma$, between $\overline{AB}$ and $\overline{AD}$, to achieve the intended $\pi/2$ relative phase, depends on  the refractive index of the material
and on the number of internal reflection, $N_{_{\mathrm{ref}}}$. For $N_{_{\mathrm{ref}}}=4$, by using
\begin{equation}
\gamma=\arctan\left[\,\sqrt{\frac{n^{\2}-1 \pm \sqrt{(n^{\2}-1)^{^{2}}-4\,n^{\2}\tan^{\2}(\pi/4\,N_{_{\mathrm{ref}}})}}{2\,n^{\2}\tan^{\2}(\pi/4\,N_{_{\mathrm{ref}}})}} \,\right]\,\,,
\end{equation}
we find $\gamma=43.08^{\mathrm{o}}$ ($\overline{AD}\approx 2.5\,\overline{AB}$) and $74.51^{\mathrm{o}}$  ($\overline{AD}\approx 14\,\overline{AB}$) for acrylic and $\gamma=50.71^{\mathrm{o}}$ ($\overline{AD}\approx 3.5\,\overline{AB}$) and $72.06^{\mathrm{o}}$  ($\overline{AD}\approx 12\,\overline{AB}$) for water. To obtain the effect of an half wave plate, two rhombs can be used in tandem (usually cemented to avoid reflections at their interfaces). It is important to observe that the rhomb angle depends on the refractive index, which is function
of the wavelength of the incoming beam. This means that for different $\lambda$ we have different rhomb angles. This clearly shows the advantage in using a block of $15$\,cm for acrylic and $12$\,cm for water and calculating the incidence angle for which we find a relative phase of $\pi/2$ and $\pi$ for different wavelengths (and consequently different refractive indexes).

\subsection{Power oscillations}

Let us now analyse, the effect of the second polariser on the transmitted beam. By choosing $\alpha=\pi/4$ and observing that $\delta_{_{\mathrm{GH}}}\ll 1$,  the relative power can be rewritten as follows
\begin{equation}
\label{relpow2}
\mathcal{P}_{\rel}\left(\mbox{$\frac{\pi}{4}$},\beta;\tau\right) = \frac{\tau^{\2}\cos^{\2}\beta \,+\,
\sin^{\2}\beta \,+\,\tau\,\sin2 \beta \,\cos\Phi_{_{\mathrm{GH}}}}{\tau^{\2} +1}\,\,.
\end{equation}
For incidence angles for which $\tau\approx 1$, circularly polarised light ($\Phi_{_{\mathrm{GH}}}=\pi/2$) guarantees $\mathcal{P}_{\rel}=1/2$ independently of the angle used in the second polariser. This is shown in Figs.\,3 (acrylic) and 4 (water) where the relative power is plotted as a function of the incidence angles for different values of $\beta$. It is also interesting to calculate the mean power  between two complementary angles, i.e. $\beta$ and $\pi-\beta$,
\begin{equation}
\label{delta}
\rho(\beta,\tau) =\frac{
\mathcal{P}_{\rel}\left(\mbox{$\frac{\pi}{4}$},\beta;\tau\right) + \mathcal{P}_{\rel}\left(\mbox{$\frac{\pi}{4}$},\pi-\beta;\tau\right)}{2} =
 \frac{\tau^{\2}\cos^{\2}\beta \,+\,
\sin^{\2}\beta}{\tau^{\2} +1}\,\,.
\end{equation}
By increasing the incidence angle, $\tau$ becomes different from 1 and  the complementary powers, if  $\beta\neq\pi/4$, are  no longer specular functions with respect to $1/2$. This breaking of symmetry is clearly due to the difference existing between the TM and the TE transmission coefficient. The choice of $\beta=\pi/4$ not only preserves the symmetry but also maximises the amplitude of the power oscillation.  By setting $\alpha=\beta=\pi/4$ and using $\delta_{_{\mathrm{GH}}}\ll 1$ ,  the relative power (\ref{relpow}) becomes
\begin{equation}
\label{max}
\mathcal{P}_{\rel}(\mbox{$\frac{\pi}{4}$},\mbox{$\frac{\pi}{4}$};\tau) \,= \,\left(\,1 \,+\, \frac{2\,\tau}{1+\tau^{\2}}\, \cos\Phi_{_{\mathrm{GH}}}\right)\,/\,2\,\,.
\end{equation}
In view of possible experimental implementations, it is interesting to present the oscillation power  in terms of the dielectric block length instead of their number. Given the $xz$ planar dimensions of the unitary block, the incidence angle, for which we find two internal reflections and for which the beam exiting point is at the same height of the incoming one, can be obtained from Eq.\,(\ref{eq:LAB}),
\begin{equation}
\sin\theta = n\,\,\mbox{\huge $/$}\,\sqrt{1 + \left(\frac{\overline{AD}}{2\,\overline{AB}} \right)^{\2}}\,\,.
\end{equation}
Fixing  $\overline{AB}=1.0\,\mathrm{cm}$, for a dielectric block of length $\overline{AD}=2.5\,\mathrm{cm}$, the incidence angle is $68.56^{\mathrm{o}}$ (acrylic) or  $56.31^{\mathrm{o}}$ (water).  For a dielectric block of length $5.0\,\mathrm{cm}$, we find $33.60^{\mathrm{o}}$ (acrylic) and  $29.62^{\mathrm{o}}$ (water), and, finally, for a dielectric block of length  $10.0\,\mathrm{cm}$, we have $16.99^{\mathrm{o}}$ (acrylic) and  $15.14^{\mathrm{o}}$ (water). From Fig.\,2, for the acrylic case and incidence at $68.56^{\mathrm{o}}$ we find the minimal relative power after 2 blocks of $2.5$\,cm and the maximal one after 4 blocks. This is clearly visible in Fig.\,5(e) where the relative power is plotted as a function of the dielectric block length.  For incidence at   $33.60^{\mathrm{o}}$, the minimum is found after 3 unitary blocks of $5$\,cm. This is also confirmed by the plot in Fig.\,5(c).

The choice of $2.5$\,cm as unitary length was made to optimise the power measurements. For example, for a unitary  acrylic dielectric block of $2.5$\,cm, we can carry out  a measurement for incidence at $68.56^{\mathrm{o}}$ (where we have two internal reflections) but also for $33.60^{\mathrm{o}}$ (where we have only a single internal reflection). For two dielectric blocks, we have the possibility to obtain 3 experimental measurements, 4 internal reflections for   $68.56^{\mathrm{o}}$, 2 internal reflections for $33.60^{\mathrm{o}}$, and, finally a single internal reflection for $16.99^{\mathrm{o}}$. In Fig.\,5, circular dots represent an even number of internal reflections,  whereas the triangular dots represent an odd number of internal reflections. This distinction was only made to recall that for these measurement the camera has to be moved to the source side. Indeed, for an odd number of internal reflections the outgoing beam is no longer parallel to the incident one, but undergoes a specular reflection.

\section{Conclusions}

Optical phases and optical paths are intrinsically related. This allows interferometers to access optical path related information via optical phase interference, as well as to induce phase shifts by means of changes in the relative length of split paths. The original Michelson interferometer, for example, was devised to detect a directional dependent difference in the refractive index of vacuum along two orthogonal paths, while the Mach-Zehnder interferometer uses interference patterns to detect differences in optical paths' lengths. This relation is the reason why the distance travelled by split beams is usually a directly controllable parameter in most interferometers. In the present work, we have proposed a different approach, that is, to use as a beam splitter, i.e. a system that responds in an intrinsically different manner to different polarisation states.
The experimental set-up of this GH interferometer is analogous to an ellipsometer set-up, but while ellipsometers are devoted to the characterisation of thin films, the GH interferometer evaluates the phase component in the GH phenomenon. The fundamental principle behind both techniques is the possibility to force interaction between phases associated to orthogonal directions with the use of a polariser.

We showed as the relative GH phase induces a power oscillation that can be detected through experiments by an appropriate choice of incidence angles and unitary blocks. For example for acrylic and water, it is possible, by using unitary blocks of $2.5$\,cm and fixed incidence angles, to reproduce full cycles of oscillations as depicted in Fig.\,5. The relative power oscillation is also an indirect measurement of the breaking of circular polarisation. Indeed, circular polarisation is obtained for a relative GH phase $\phi_{_{\mathrm{GH}}}=\pi/2$ which implies $\mathcal{P}_{\rel}(\frac{\pi}{4},\frac{\pi};\tau)=1/2$.

In looking for the power oscillations, we also gained as a bonus the possibility to build simple parallelepiped dielectric blocks which, for different incidence angles, reproduce the effect of quarter and half wave plates.
The advantage to use the new dielectric configuration as retarder is that it can be used for different wavelengths.
The refractive index depends on the wavelength of the incidence beam. For different $\lambda$, by using the standard Fresnel rhombs we have to build different dielectric blocks with different $\gamma$ angles. In our proposal, we have to change the incidence angle, but  leaving unaltered the geometry of the dielectric block.

Finally, we also notice that, as the optical path is a function of the incidence angle and of the refractive index, the interference pattern of the transmitted power is highly sensible to the optical system and could thus be used to sense rotation, test optical components and control polarisation.


\ColumnFigure{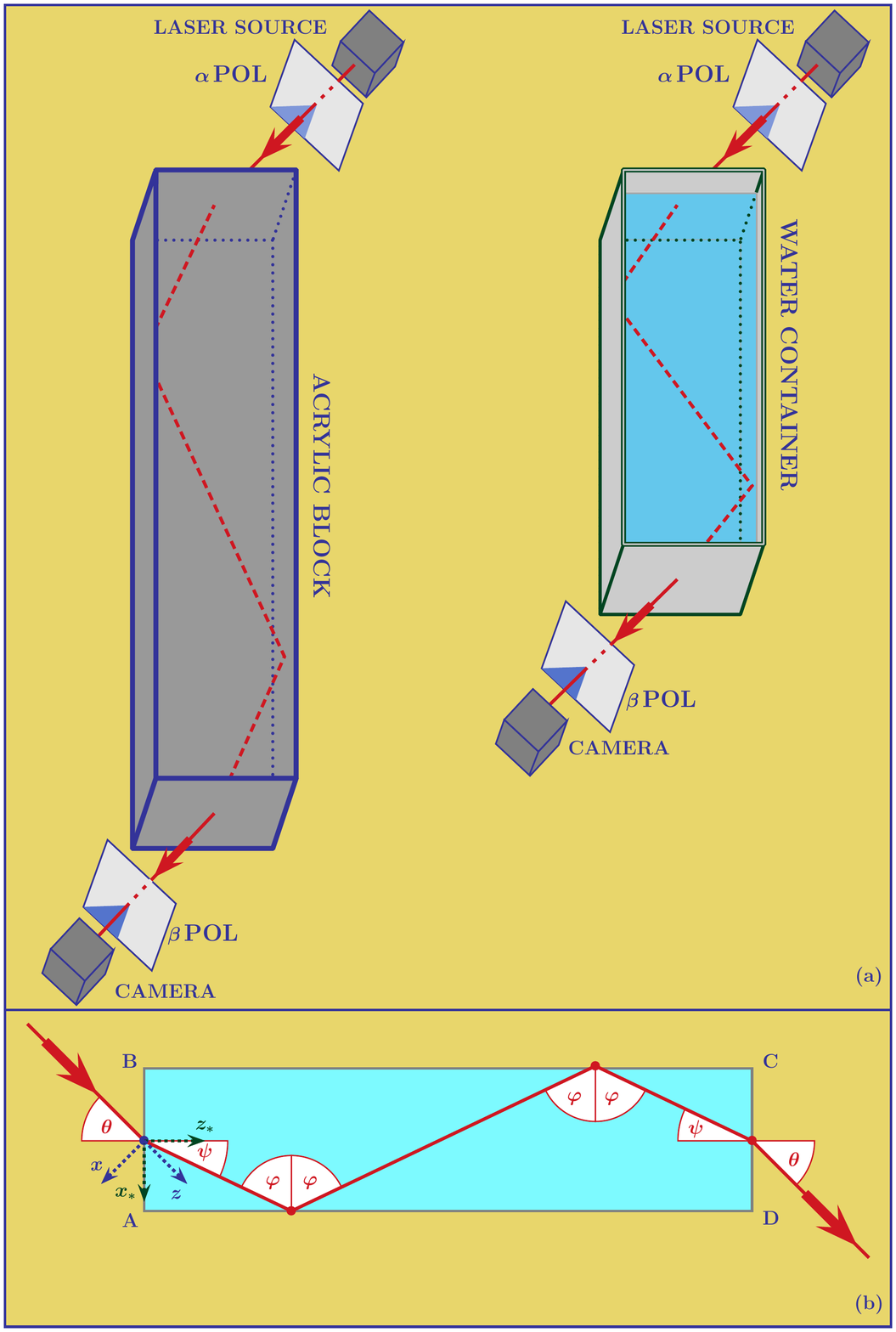}{The 3D schematic representation of the experimental setup and the planar geometry of the dielectric block are respectively sketched in (a) and (b). In (b), $z$ is the laser's propagation direction, $z_*$ the normal to the left and right sides of the dielectric block and $x_*$ the normal to the upper and lower ones. The refracted angle $\psi$ is determined by the Snell law and the internal reflection angle by the geometry of the system.}
\ColumnFigure{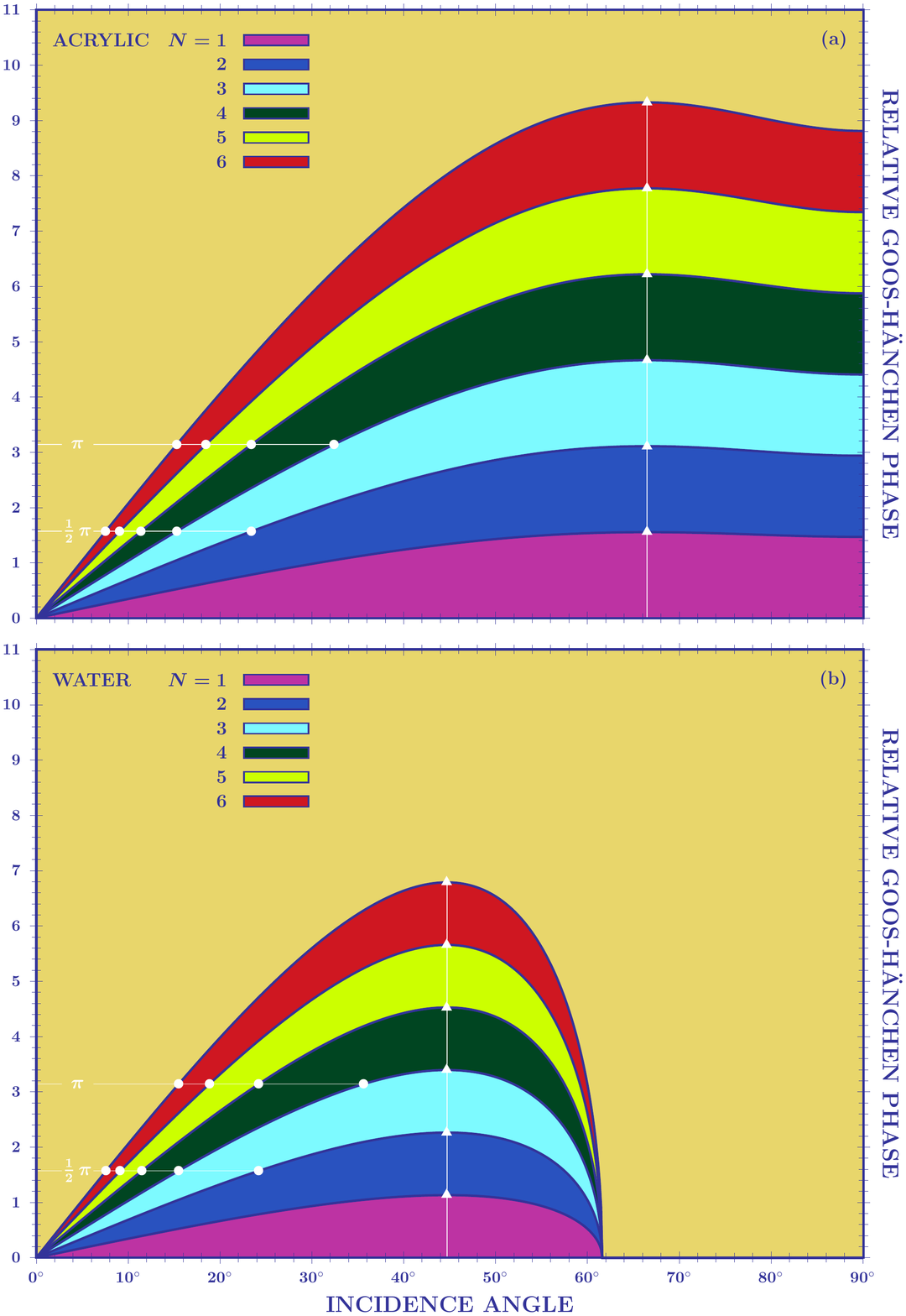}{The relative GH phase is plotted as a function of the incidence angle for a dielectric composed of different number of unitary blocks in (a) for acrylic and in (b) for water. The maximum is found at $66.52^{\mathrm o}$ for acrylic and $44.72^{\mathrm o}$ for water. In the figures, we also find the incidence angle for which it is possible to simulate quarter ($\pi/2$) and half ($\pi$) wave plates for acrylic (a) and water (b).}
\ColumnFigure{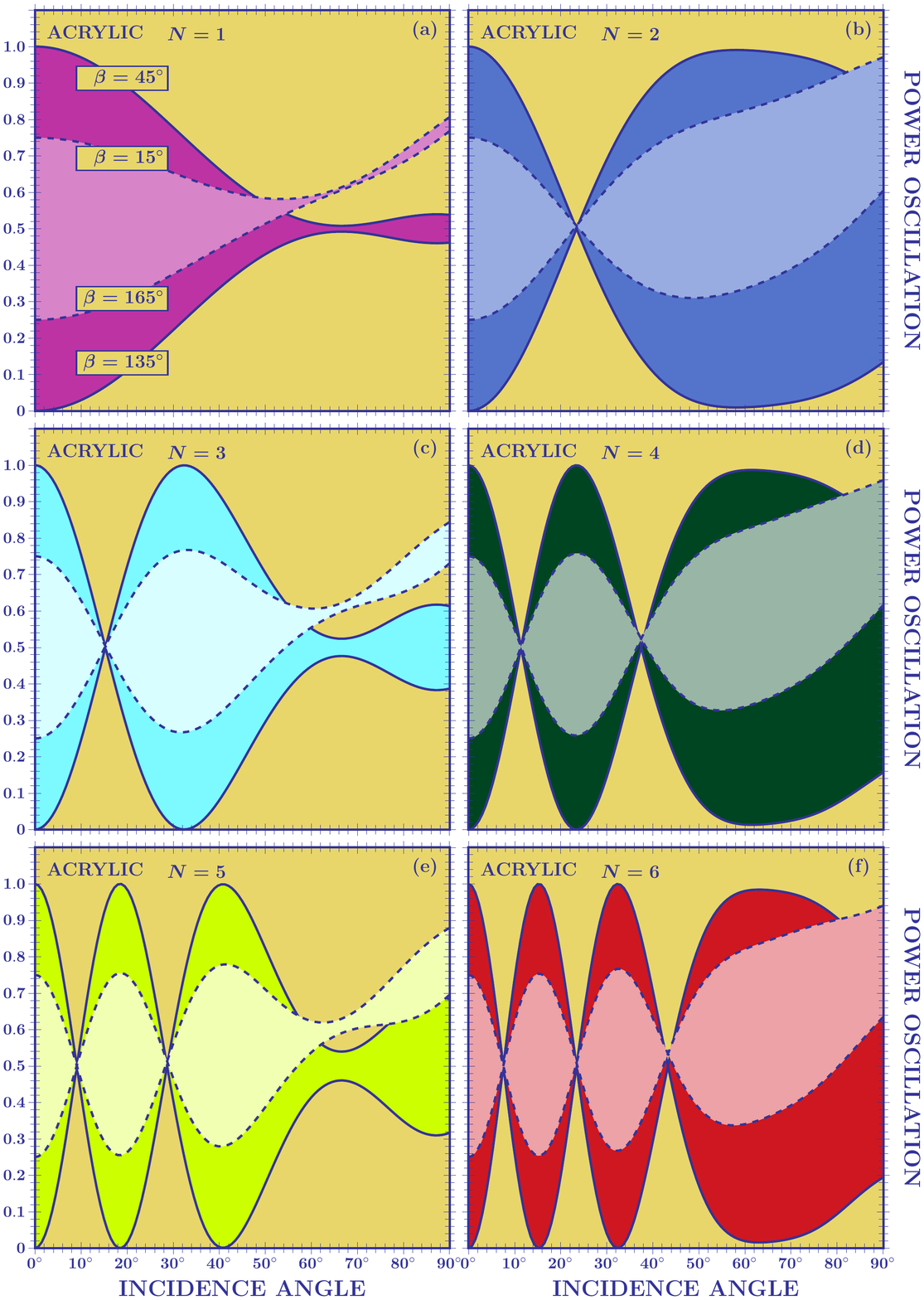}{The acrylic power oscillation, for different number of unitary blocks and $\alpha=\pi/4$, is plotted as a function of the incidence angle for $\beta=\pi/12$ (dashed line), $\pi/4$ (continuous line) and their complementary angles. The curves for $\beta=\pi/4$ and $3\pi/4$ are symmetric with respect to $1/2$. For a different choice of complementary angles, this symmetry is broken as the incidence angle increases.}
\ColumnFigure{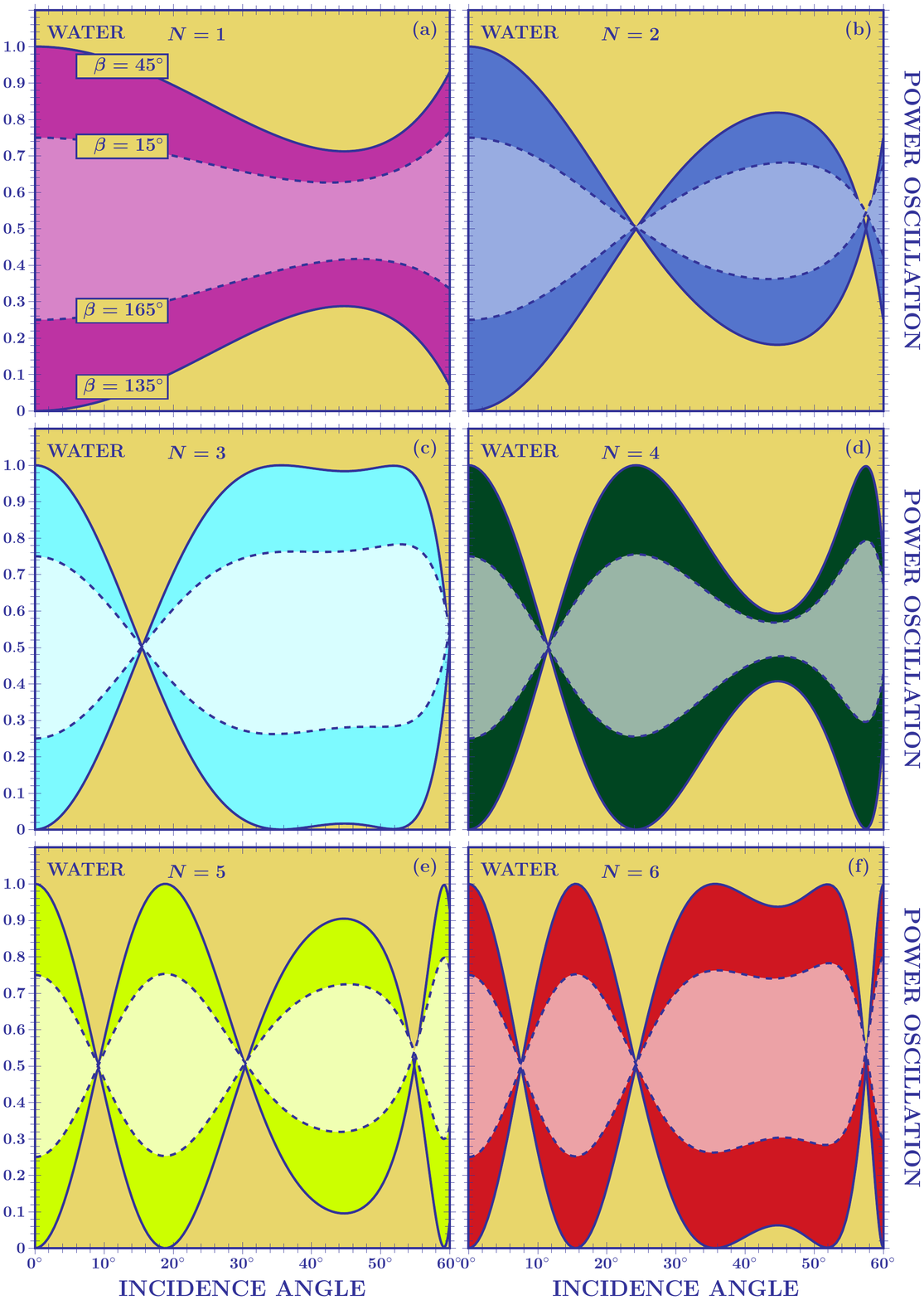}{The water power oscillation, for a different number of unitary blocks and $\alpha=\pi/4$, is plotted as a function of the incidence angle for $\beta=\pi/12$ (dashed line), $\pi/4$ (continuous line) and their complementary angles.
In this case the breaking of symmetry is less evident in comparison to the acrylic case, as we have an upper limit, $61.63^{\mathrm o}$, for the incidence angle.}
\ColumnFigure{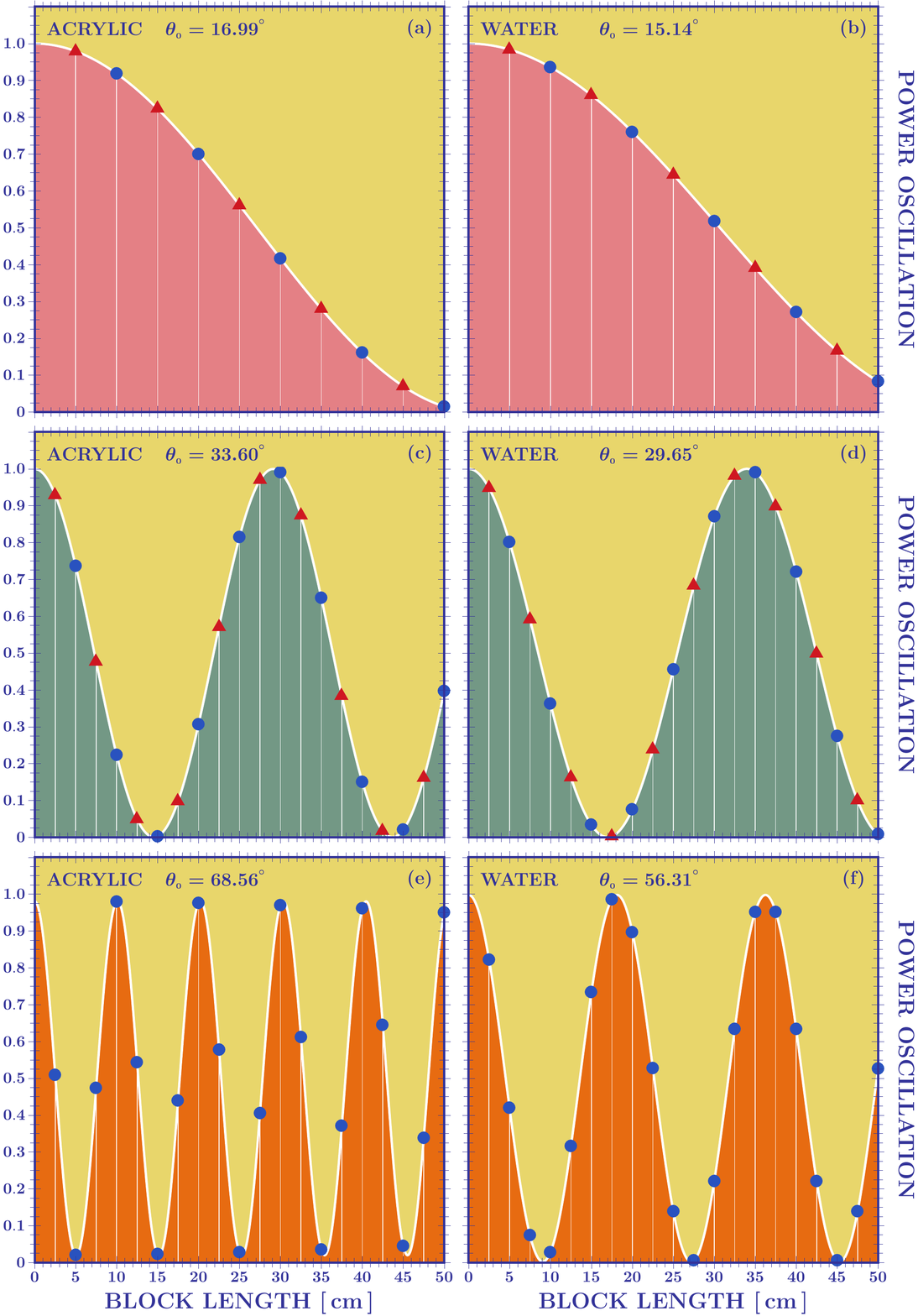}{Power oscillation as a function of the dielectric block length for acrylic and water. The incidence angles have been chosen to give for the same length more power oscillation measurements. By increasing the incidence angle we increase the number of oscillations. The {\large ${\color{blue} \bullet}$} and ${\color{red} \blacktriangle}$ dots represent the experimental measurements done with the camera respectively positioned to lower (see Fig.\,1) and  upper side of the dielectric block.}

\end{document}